\documentclass[usenatbib]{basi}
\usepackage[T1]{fontenc}
\usepackage[british]{babel}
\usepackage[varg]{txfonts}
\usepackage{rotating}
\usepackage{dcolumn}
\begin{document}

\title[Recurrent novae as progenitors of SNe Ia]{Recurrent
novae as progenitors of Type Ia supernovae}

\author[M.\ Kato  \&  I.\ Hachisu]
{ Mariko Kato$^1$\thanks{email: \texttt{mariko@educ.cc.keio.ac.jp}} and 
Izumi Hachisu$^2$  \\
    $^1$ Keio University, Hiyoshi, Kouhoku-ku, Yokohama, 223-8521, Japan\\ 
    $^2$ Dept. of Earth Science and Astronomy, College of Arts and
    Sciences, The University of Tokyo, Komaba,\\
    Meguro-ku, Tokyo 153-8902, Japan}

\pubyear{2012}
\volume{40}
\pagerange{\pageref{firstpage}--\pageref{lastpage}}

\date{Received 2012 September 05; accepted 2012 October 20}

\setcounter{page}{393}
\maketitle

\label{firstpage}

\begin{abstract}
Recurrent novae are binaries harboring a very massive white dwarf (WD), 
as massive as the Chandrasekhar mass, because of their short recurrence
periods of nova outbursts of 10--100 years.  Thus, recurrent novae
are considered as candidates of progenitors of Type Ia supernovae (SNe~Ia).
In fact, the SN~Ia PTF~11kx showed evidence that its progenitor is
a symbiotic recurrent nova.
The binary parameters of recurrent novae have been well determined, 
especially for the ones with frequent outbursts, U Sco and RS Oph, 
which provide useful information on the elementary processes
in binary evolution toward SNe~Ia. 
Therefore we use them as testbeds for binary evolution models. 
For example, the original double degenerate (DD) scenario cannot
reproduce RS Oph type recurrent novae,
whereas the new single degenerate (SD) scenario
proposed by Hachisu et al. (1999) naturally can. 
We review main differences between the SD and DD scenarios, 
especially for their basic processes of binary evolution.  
We also discuss observational support for each physical process. 
The original DD scenario is based on the physics in 1980s, 
whereas the SD scenario on more recent physics including the new opacity, 
mass-growth efficiency of WDs, and optically thick winds developed
in nova outbursts.  
\end{abstract}


\section{Introduction}\label{sec:intro}

Nova outbursts are a thermonuclear runaway event on a white dwarf (WD). 
After the unstable hydrogen shell-burning sets in, the envelope greatly 
expands and the WD becomes very bright to reach optical maximum. 
Structures of such expanded envelopes are subject to the opacity
that regulates the energy flux.
In the beginning of 1990s opacity tables were revised by the OPAL
\citep{igl87,rog92,igl96} and OP \citep{sea94} projects.
These new opacities 
showed a prominent peak at a temperature of $\log T$ (K) $\sim 5.2$ due 
to iron lines, and brought drastic changes into stellar structures, 
especially when the luminosity is close to the Eddington luminosity. 
Using the new opacities, people solved many long-standing problems
in pulsation, stellar evolution, and nova outbursts
\citep[e.g.][and references therein]{bre93,scha92,kat94h,pri95,gau96}.  
Now a days the OPAL opacities are widely used in astronomy. 

Nova theory has been greatly influenced by the new opacities.
The strong peak in the opacity accelerates winds during nova outbursts and, 
as a result, timescales of nova outbursts are drastically shortened. 
With the old opacities \citep[the Los Alamos opacities, 
e.g.,][]{cox70a,cox70b,cox76}, we had to assume
that novae occur only in very massive WDs ($M_{\rm WD}\gtrsim1.3~M_\odot$)
or that the frictional effect by a companion star to the WD is very
effective in ejecting most of the WD envelope in a dynamical timescale.
These assumptions were required for a relatively short duration
of nova outbursts ($\sim 1$ yr).  With the new opacities,
however, these assumptions are not needed anymore.  In and after 1990
many numerical calculations were done with the OPAL opacities.
We have now strong accelerations of winds that blow most of the envelope
mass in a relatively short timescale of $\sim1$~yr.  As a result,
calculated nova timescales are drastically shortened \citep{kat94h,pri95}.
Kato developed an optically thick wind theory to follow the extended stage
of nova outbursts and her group succeeded in reproducing nova light curves
from slow to very fast novae \citep[e.g.,][]{hac06kb,hac10k}.  
Now we understood basic properties of novae,  
such as duration, expansion speed, multiwavelength properties, 
and light curves for all speed classes of novae. 

Type Ia supernovae (SNe~Ia) are characterized in principle
by spectra with strong Si~{\scriptsize II} but no hydrogen lines
at maximum light.  SNe Ia play very important roles in astrophysics
as a standard candle for measuring cosmological distances and
as main producers of iron group elements in chemical evolution
of galaxies.  It is commonly agreed that the exploding star is
a mass-accreting carbon-oxygen (C+O) WD.  For example, the SN Ia 2011fe
indicates its size of the exploding star
as small as $R < 0.02~R_\odot$ \citep{blo12}, consistent with its
WD origin.  However, it is not clarified yet whether
the WD accretes H/He-rich matter from
its binary companion (the so-called single degenerate (SD) scenario),
or two C+O WDs merge (the so-called double degenerate (DD) scenario)
\citep[e.g.,][for review]{hil00, nom00}.  It was before the advent of
the OPAL opacity that \citet{web84} and \citet{ibe84} proposed the DD
scenario.  In this DD scenario, 
intermediate-mass (3--8~$M_\odot$) binaries undergo the first and
second common envelope phases and finally become a double WD system
with a very short orbital period of $\lesssim 3$~hr.  They can merge
within a Hubble time due to orbital energy and angular momentum losses by
gravitational radiation.  If the total mass of a merged object exceeds
the Chandrasekhar mass, it could explode as an SN Ia.
This is the traditional DD scenario.

\begin{figure}
\begin{center}
\begin{tabular}{p{6.5cm}cp{6cm}}
\raisebox{-\height}{\includegraphics[width=7cm]{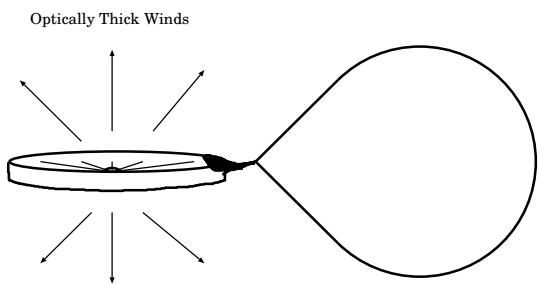}} & \quad &
\caption{
Accretion wind evolution in the SD scenario.  Optically thick winds blow 
from a mass-accreting WD when the mass-transfer rate from its lobe-filling 
companion exceeds a critical rate, $\dot M_{\rm cr}$.
Hydrogen steadily burns on the WD at the same rate of $\dot M_{\rm cr}$
(typically $\sim1\times10^{-6}~M_\odot$ yr$^{-1}$ in massive WDs), while 
transferred matter exceeding this burning rate is blown in the wind
as shown in the figure.
\label{fig:accretionwind}}
\end{tabular}
\end{center}
\end{figure}

\citet{hkn96} proposed a new idea, dubbed ``accretion wind evolution,''
as a new elementary process in the binary evolution that leads to SN Ia
explosions.  This accretion wind evolution
is based on the optically-thick nova wind theory,
illustrated in Fig. \ref{fig:accretionwind}.
The WD accretes matter from the accretion disk and, at the same time, 
blows optically thick winds.  The wind blows as far as the WD accretes
matter from the equator at sufficiently high rates.  
In the accretion wind evolution, the strong and fast winds
carry mass and angular momentum, so the mass transfer can be stabilized.
This means that the so-called second common envelope evolution
does not occur when the accretion wind evolution is realized.
This changed the traditional binary evolution scenarios.
Thus, \citet{hkn96} established a new progenitor system of SNe Ia
which consists of a mass-accreting WD and a red-giant (RG) donor star.
In the accretion wind evolution, WDs can grow in mass and 
explode as an SN Ia.  We call this the ``accretion-wind single degenerate
(awSD) scenario,'' if it is necessary to distinguish it from the old SD
scenario first proposed by \citet{whe73}.

\citet{li97} adopted this idea as a fundamental process of binary
evolution and found that, in some binaries,
the second common envelope evolution does not  
occur and the WD can grow in mass up to the Chandrasekhar limit.
These binaries consist mainly of a WD and a main-sequence (MS) star
at the stage of mass transfer (WD+MS systems).  This is also a new way
that leads to SN Ia explosions and dubbed ``MS channel.''
After that, \citet{hkn99} and \citet{hknu99} reformulated basic
processes of binary evolution and found that the WD+RG systems 
(the so-called ``symbiotic channel'') contribute to the SN Ia rate
as well as the WD+MS systems.

Recurrent novae are characterized by multiple recorded outbursts
of novae.  So far, ten objects were registered in our Galaxy as listed
in Table \ref{table.RN}.  They should have a very massive WD close to
the Chandrasekhar mass because of the short recurrence time of outbursts
\citep[e.g.][see also Fig. \ref{fig:MwdMdot}]{hac01b, pri95}.
The present evolutionary status of
these recurrent novae can be well understood from the theoretical
point of view, if we take into account the accretion wind evolution.
The binary parameters of several recurrent novae locate just in
the binary parameter regions of the MS and symbiotic channels of SNe Ia
\citep[e.g.,][see also Fig. \ref{fig:SD}]{hac01b, hkn12b}.
Thus we use the binary parameters of recurrent novae as testbeds
of scenarios on Type Ia supernova evolution.

In the original DD scenario proposed by \citet{web84}
and \citet{ibe84}, the accretion wind evolution was not included,
so the binary could evolve to a pair of WDs
after the second common envelope evolution. 
Thus, the first essential difference between the SD and DD scenarios is
in the accretion wind evolution, which was based on the optically
thick wind theory that was developed after the new opacities
appeared in 1990s.
In the DD scenario, binaries undergo the second common envelope evolution
and the evolutionary path is essentially governed by the
parameters of common envelope evolution \citep[e.g.,][]{web08}. 
The DD scenario is also based on other different assumptions from the
SD scenario, some of which are based on 1980s physics.  In the present
review, we critically examine these assumptions in light of observations
of recurrent novae and other related objects.

Section \ref{sec:RNobs} introduces observational properties of
Galactic recurrent novae.  
Section \ref{sec:nova} summarizes   
our theoretical understanding of novae, especially some issues closely
related to SN Ia progenitors, i.e., stability of accreting WDs
and how to determine the WD mass.  Section \ref{sec:SD} shows that 
these recurrent novae locate at the final stage of evolutionary path to SNe Ia 
in the SD scenario.  Section \ref{sec:elementary} examines several 
elementary processes that are important in binary evolution. 
Section \ref{sec:relatedobj} deals with some objects closely related
to the SN~Ia progenitor scenarios.

\newcolumntype{d}[1]{D{.}{.}{.}{.}{.}{#1}}
\begin{table}
  \caption{Recurrent novae}
  \medskip
  \begin{center}
    \begin{tabular}{llrcccc}\hline
      object $^\mathrm{a}$& 
        \multicolumn{1}{c}{outburst year}  & 
        \multicolumn{1}{c}{$P_{\rm rec}$ $^\mathrm{b}$}  & 
        \multicolumn{1}{c}{$P_{\rm orb}$}  &
        \multicolumn{1}{c}{$t_3$ $^\mathrm{c}$}  \\ 
        \multicolumn{1}{c}{}  & 
        \multicolumn{1}{c}{}  & 
        \multicolumn{1}{c}{(yr)}  & 
        \multicolumn{1}{c}{(day)} &
        \multicolumn{1}{c}{(day)}  \\\hline
CI Aql   &1917,1941,2000  & 24 & 0.618 &  32   \\
V394 CrA  & 1949,1987  &     38 &  1.52 & 5.2  \\
V2487 Oph & 1900,1998 & 98 &$\sim 1$  & 8 \\
U Sco &1863,1906,1936,1945 & 8 &1.23 & 2.6 \\
 &1979,1987,1999,2010 &  &  & \\
T CrB    &1866,1946       & 80 & 228 & 6   \\
RS Oph &1898,1933,1958,1967 & 9 & 453.6 & 14 \\
 &1985,2006 & & & \\
V745 Sco    & 1937,1989&  52 & 510 & 9  \\
V3890 Sgr     &1962,1990 & 28 & 520 & 14  \\
IM Nor    &1920,2002      & 82  &0.103 & 80 \\
T Pyx & 1890,1902,1920,1944 & 12 & 0.076 & 62 \\
 & 1966, 2011 &  \\
     \hline
    \end{tabular}\\[5pt]
    \begin{minipage}{13cm}
  \small Notes:  (a) references for each objects are as follows.   
CI Aql: outburst year \citep{col09}, $P_{\rm orb}$ \citep{man95}.
V394 CrB: $P_{\rm orb}$ \citep{sch09}.
V2487 Oph: outburst year \citep[][who suggest the recurrence period to 
be possible 18 yr instead of 98 yr from discovery frequency]{pag09}, 
$P_{\rm orb}$ \citep{schae10}.
U Sco: \citet{schae10} suggested 1917 and 1969 eruptions, $P_{\rm orb}$
\citep{schae10}. 
T CrB: $P_{\rm orb}$ \citep{lin88}.
RS Oph: \citet{schae10} suggested 1907 and 1945 outburst, $P_{\rm orb}$
\citep{bra09}. 
V745 Sco: $P_{\rm orb}$ \citep{sch09}.
V3890 Sgr: $P_{\rm orb}$ \citep{sch09}.
IM Nor: $P_{\rm orb}$ \citep{wou03}.
T Pyx: $P_{\rm orb}$ \citep{schae92}. \\
(b) shortest recurrence period. \\
(c) all $t_3$ are taken from \citet{schae10}. 
    \end{minipage}
  \end{center}
\label{table.RN}
\end{table}

\newcolumntype{d}[1]{D{.}{#1}}
\begin{table}
  \caption{WD mass in recurrent novae and related objects}
  \medskip
  \begin{center}
    \begin{tabular}{llrrlccl}\hline
      object $^\mathrm{a}$& 
        \multicolumn{1}{c}{WD mass}  & 
        \multicolumn{1}{c}{$t_2$ $^\mathrm{b}$} &
        \multicolumn{1}{c}{distance} &
        \multicolumn{1}{c}{$E(B-V)$ $^\mathrm{c}$} &
        \multicolumn{1}{c}{$m_{V,\rm max}$} &
        \multicolumn{1}{c}{$M_{V,\rm max}$} &
        \multicolumn{1}{c}{$\Delta M_{V,\rm MMRD}$} \\
        &
        \multicolumn{1}{c}{($M_\odot$)}  & 
        \multicolumn{1}{c}{(day)}   &
        \multicolumn{1}{c}{(kpc)}   &
        \multicolumn{1}{c}{}  &
        \multicolumn{1}{c}{}  &
        \multicolumn{1}{c}{} &
        \multicolumn{1}{c}{(1),~\ (2)} \\
\hline
CI Aql    & $\sim1.2$ & 25.4 & 4.7 & 0.85 & 8.8 & $-7.2$ & $-0.3,-0.4$ \\
V394 CrA  & $\gtrsim1.37$ & 2.4 &  & 0.25 & 7.2 & & \\
V2487 Oph & $\gtrsim1.37$ & 6.2 & & 0.76$^\mathrm{d}$ & 9.5 & & \\
U Sco$^*$   & $\gtrsim1.37$ & 1.2 & 6.7 & 0.35$^\mathrm{d}$ & 7.7 & $-7.5$ & $-3.0^*,-1.5$ \\
T CrB$^*$    & $\gtrsim1.35$ & 4.0 & 0.96 & 0.065 & 2.5 & $-7.6$ & $-2.4^*,-1.3$ \\
RS Oph   & $\sim1.35$ & 6.8 & 1.6 & 0.75$^\mathrm{d}$ & 4.8 & $-8.5$ & $-0.9,-0.3$ \\
V745 Sco  & $\gtrsim1.35$ & 6.2 & & 0.82$^\mathrm{d}$ & 9.4 & & \\
V3890 Sgr  & $\gtrsim1.35$ & 6.4 & & 0.56$^\mathrm{d}$ & 8.1 & \\
%
IM Nor &  & 50.0 & 3.4 & 0.80 & 8.5 & $-6.7$ & $-0.3,-0.4$ \\
T Pyx & & 34.6 & 3.2 & 0.25 & 6.3 & $-7.0$ & $-0.2,-0.3$ \\
\hline
V407 Cyg$^*$ & $\gtrsim1.37$ & 5.9 & 2.7 & 0.50 & 7.1 & $-6.6$ & $-3.0^*,-2.3$ \\
V838 Her & $\sim1.35$ & 1.4 & 4.0 & 0.43$^\mathrm{d}$ & 5.4 & $-8.9$ & $-1.6,-0.1$ \\
V2491 Cyg & $\sim1.3$ & 4.8 & 13.0 & 0.23 & 7.5 & $-8.8$ & $-0.9,-0.1$ \\
     \hline
    \end{tabular}\\[5pt]
    \begin{minipage}{13cm}
  \small Notes:  (a) references for each objects are as follows.   
CI Aql:  WD mass \citep{hac01a}, distance \citep{hac12kd}. 
V394 CrB: WD mass \citep{hac00}.
U Sco: WD mass \citep{hkkm00} but a super-Chandrasekhar mass  WD
is suggested by \citet{hksn12a}, distance \citep{hkkm00, hkkmn00}.
T CrB: WD mass \citep{hac01b}, distance and $E(B-V)$ \citep{bel98}.
RS Oph:   WD mass \citep{hac06a,hac06kb}, distance \citep{hje86}.
V745 Sco: WD mass \citep{hac01b}.
V3890 Sgr: WD mass \citep{hac01b}.
T Pyx: distance \citep{hac12kd}.
V407 Cyg: WD mass \citep{hac12kc}, $t_2$ and distance \citep{mun12},
$E(B-V)$ \citep{sho11}.
V838 Her: WD mass \citep{kat09v838her}, $t_2$ \citep{har94},
distance is calculated
from the results of \citet{kat09v838her} and the present value of $E(B-V)$.
V2491 Cyg:  WD mass \citep{hac09}, $t_2$ \citep{mun11},
distance is calculated from
the same method in \citet{hac12kd}, $E(B-V)$ \citep{mun11}. \\
(b) all $t_2$ for recurrent novae are taken from 
\citet{schae10} unless otherwise specified. \\ 
(c) all $E(B-V)$ for recurrent novae are taken from \citet{schae10} unless
otherwise specified. \\
(d) $E(B-V)$ is obtained from the Galactic dust absorption map of
NASA/IPAC (http:// irsa.ipac.caltech.edu/ applications/DUST/).
\label{tab:WDmass}
    \end{minipage}
  \end{center}
\end{table}


\section{Recurrent novae: observations}\label{sec:RNobs}

Recurrent novae are a small subgroup of novae that have multiple
recorded outbursts.  Their characteristic properties are summarized
as follows.\\
\vspace{0.3cm}\\
\ \ \ 1. Recurrence periods between outbursts are as short as $\sim10$--100~yrs.\\
\ \ \ 2. Optical light curves show a rapid decline.\\
\ \ \ 3. Optical light curves often show a mid-plateau phase. \\
\ \ \ 4. In the later phase of outbursts,
it often becomes a transient supersoft X-ray source (SSS).\\
\ \ \ \ \ \ \ This SSS period almost overlaps with the mid-plateau phase.\\
\ \ \ 5. No heavy element enhancement is detected in its ejecta. \\

Only about ten recurrent novae are known in our Galaxy, but this does not
directly suggest a small population of recurrent novae.
Recurrent nova eruptions develop so fast and their detections are not
easy especially in their short bright phases.
There are potentially a number of recurrent novae that are not yet
identified as a recurrent nova \citep{pag09}.
All of the recurrent novae are now regarded as a thermonuclear runaway event
owing to instability of hydrogen nuclear burning in a geometrically
thin shell on a WD.  \citet{web87} categorized recurrent novae
into three subclasses,
in which T CrB and RS Oph were modeled as a pair of an MS and an RG,
but now it is clear that they are a binary consisting of
a WD and an RG \citep{sel92, web08}.
Also a disk instability model of RS Oph outbursts \citep{kin09,ale11} 
is incompatible with many observational indications toward
a thermonuclear runaway event \citep[e.g.,][]{nels11}.

The recurrent novae are divided into three groups 
depending on the type of companion star:\\
\vspace{0.3cm}\\
\ \ \ 1. Slightly evolved MS stars: CI Aql, V394 CrA, V2487 Oph, and U Sco.\\
\ \ \ 2. Red-giant stars: T CrB, RS Oph, V745 Sco, and V3890 Sgr. \\
\ \ \ 3. Red dwarf stars: IM~Nor and T~Pyx. \\
\vspace{0.2cm}\\
Here ``slightly evolved MS'' means that the companion star is
a main-sequence star that has evolved off the zero-age main-sequence (ZAMS)
and its radius has expanded a little.  These three groups 
are sometimes called the U Sco, T CrB (or RS Oph), and T Pyx 
subclasses \citep[e.g.][]{war95}, respectively. 
Table \ref{table.RN} lists the recorded outburst year,
shortest recurrence period, orbital period, and $t_3$ time
for all the Galactic recurrent novae ever reported, where
$t_m$ ($m=2$ or 3) is the day during which a nova decays by
$m$-th magnitude from optical maximum.
Table \ref{tab:WDmass} shows 
the WD mass estimated from the light curve
analysis explained later in Section \ref{sec:nova}, $t_2$ time, 
distance $d$, extinction $E(B-V)$, apparent $V$ magnitude at maximum 
$m_{V,\rm max}$, and absolute $V$ magnitude $M_{V,\rm max}$ at maximum,
and difference between the absolute $V$ maximum observed 
and the $V$ maximum calculated from the MMRD relation.
We added three related objects in order to compare them with the
Galactic recurrent novae.  
Recurrent novae in the LMC and M31 are not listed in these Tables.

Note that in order to classify a nova into the subgroup of recurrent novae, 
just a rapid decline in the light curve, which suggests a massive WD,
is not enough.  For example, the classical nova V838 Her showed a 
light curve as steep as that of U Sco, also the light curve of V2491 Cyg
decayed similarly to that of RS Oph.
Their WD masses are estimated to be very massive, 
$1.35~M_\odot$ for V838 Her \citep{kat09v838her}, and 
$1.3~M_\odot$ for V2491 Cyg \citep{hac09}. 
However, these nova ejecta showed heavy element enrichment,
$Z=0.12$ for V838 her \citep{van97,sch07} and 
$Z=0.14$ for V2491 Cyg \citep{mun11}, suggesting a long recurrence period 
between nova outbursts during which diffusion processes work to
dredge up WD core material into hydrogen-rich envelope. 
V2491 Cyg also showed a much shorter SSS phase (10 days) than that of 
RS Oph (60 days).  A relatively long duration of a SSS phase stems
from a mass-increasing WD \citep[][see Section \ref{sec:lca}]{kat12iauR} 
while a relatively short one indicates
a mass-decreasing WD as in normal classical novae. 
Therefore, both V838 Her and V2491 Cyg are highly likely classical novae
rather than recurrent novae.

\section{Theory of Nova outbursts}\label{sec:nova}

\subsection{Stability of shell flashes}\label{sec:instability}

A mass-accreting WD has a geometrically thin envelope before a nova eruption. 
The envelope is usually cold because of radiative cooling. 
As the envelope mass increases with time, the bottom of the envelope is
gradually heated up due to compressional heating
(gravitational energy release).  
When the envelope mass reaches a critical value, the radiative cooling
almost balances with the total release of gravitational and nuclear
burning energies.  In such a situation, even a very small increase
in the temperature causes a large increase in the nuclear burning energy
release.  This increase in the energy release cannot be dissipated away
by radiation, and so results in a runaway of hydrogen nuclear burning.
This is the beginning of a nova outburst.  

\begin{figure}
\begin{center}
\centerline{\includegraphics[width=8cm]{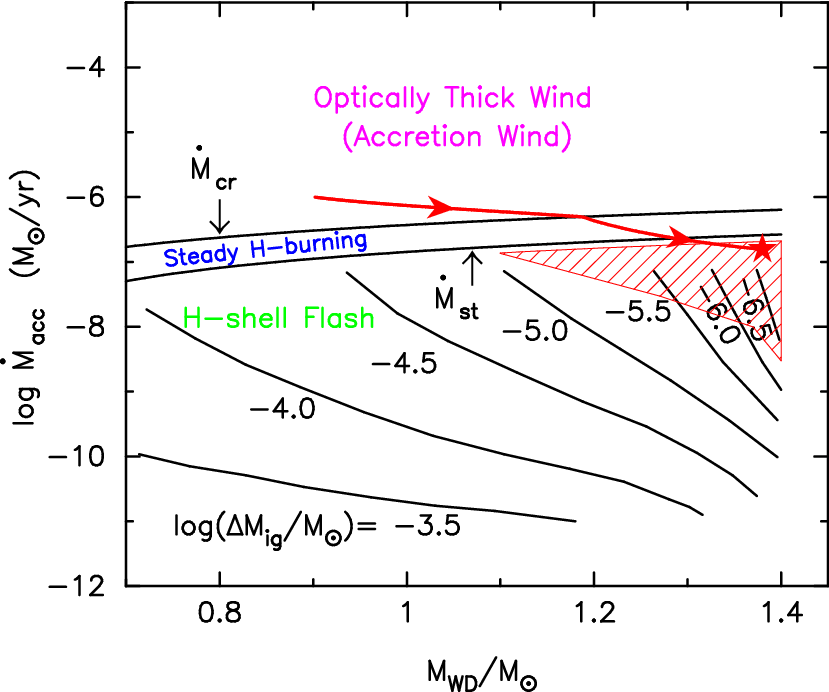}}
%
\caption{
An evolutionary path of typical SN Ia progenitors
(red solid line with arrows) on the map of WD responses
to the mass accretion rate $\dot M_{\rm acc}$.
The progenitor evolves from the optically thick wind phase, 
through SSS phase (steady H-burning phase)
and recurrent nova phase (H-shell flash phase), and
finally explodes as an SN Ia at star mark.
In the region above the line of $\dot M_{\rm cr}$
($\dot M_{\rm acc} > \dot M_{\rm cr}$), 
strong optically thick winds blow, which stabilize binary evolution
in the SD scenario.  On the other hand, a common envelope is formed
in the original DD scenario, which results in a binary consisting
of a double degenerate (WD) system. 
In the region of
$\dot M_{\rm st} \le \dot M_{\rm acc} \le \dot M_{\rm cr}$, we have
steady hydrogen shell burning with no optically thick winds.
There is no steady-state burning below the line of $\dot M_{\rm st}$  
($\dot M_{\rm acc} < \dot M_{\rm st}$).  
Instead, hydrogen shell-burning intermittently occurs and results in
a nova outburst.  The envelope mass $\Delta M_{\rm ig}$,
at which an intermittent hydrogen shell flash ignites, is also shown. 
Shadow area schematically shows the region in which the recurrence period
is shorter than 100 yr, that is, the region of recurrent novae. 
The original data are taken from \citet{nom82}, \citet{hac01a},
and \citet{hkn10}.
\label{fig:MwdMdot}}
%
%
\end{center}
\end{figure}

The ignition mass of hydrogen-rich envelope 
is obtained from the condition that the radiative cooling
is balances with the total release of gravitational and nuclear burning
energies.  This ignition mass depends on the WD mass,
mass accretion rate, WD temperature, and opacity
(i.e., chemical composition) of the envelope. 
In recurrent novae, the ignition mass does not depend on
the thermal history of the WD because the recurrence periods 
are shorter than $\sim 100$ yr. 
The ignition mass is smaller for a more massive WD because of
its large compressional heating.
Fig. \ref{fig:MwdMdot} schematically depicts ignition masses
by solid lines in the region of $\dot M_{\rm acc} < \dot M_{\rm st}$,
where hydrogen shell burning is unstable.

During nova outbursts, all of the accreted matter will be blown off.
Moreover, a part of the WD core is eroded 
due to diffusion of hydrogen into the core (see the next subsection). 
Then, the WD mass decreases after every nova outburst. 
In the case of recurrent novae,
however, hydrogen shell flash is relatively weak 
and a certain amount of processed helium is added
to the WD after every outburst.  Therefore the WD mass 
increases (see Section \ref{sec:Hflash}).  This recurrent nova
region is indicated by shadow in Fig. \ref{fig:MwdMdot}.
Their accretion rate is
as large as $\dot M_{\rm acc} \sim10^{-8}$--$10^{-7}M_\odot$~yr$^{-1}$ and
the WD is quite massive ($M_{\rm WD}\gtrsim1.1 M_\odot$).

When the mass accretion rate is larger than
$\dot M_{\rm acc}\ge\dot M_{\rm st}$,
hydrogen nuclear burning is stable and ash of hydrogen burning
accumulates on the WD.  When the mass accretion rate exceeds
$\dot M_{\rm cr}$, the envelope expands to blow
optically thick winds.  The system undergoes an accretion 
wind evolution as illustrated in Fig. \ref{fig:accretionwind} while
the binary was supposed to undergo a common envelope evolution
in the original DD scenario.  

Stability analysis of hydrogen shell-burning had been established
a long time ago based on linear analyses by \citet{si75, si80}.
His results were confirmed recently with the OPAL opacity 
\citep[][]{nom07}.  
These stability criteria were also confirmed directly
in many numerical calculations
of shell flashes \citep{pri95,jos98}. 
\citet{star04}, however, presented an opposite result that
all accreting WDs are thermally stable
(this means that no novae/recurrent novae occur),
thus, that the accreting WDs always
grow in mass to reach the Chandrasekhar mass \citep{star04}.  
Their calculations were criticized by \citet{nom07} for the reason that
their results are artifacts originating from the lack of sufficient
numerical grids. 
Recently, the same group \citep{star12} presented a completely different
result to their previous one
that all accreting WDs are unstable for shell flashes
and virtually no SSS phase exists in classical novae.
Again their results are not supported by the stability analysis
cited above as well as many observed SSS phases of classical 
novae/recurrent novae 
\citep[see, e.g.,][for an M31 nova survey]{henz10, henz11}.

For the last, we must note an inappropriate statement on
the ``proper pressure,'' which is widely used in estimating the ignition
mass of novae \citep[e.g.][]{osb11}; there is a supposed
simple relation between the pressure at the bottom of an envelope
and the ignition mass, 
$P^* = (G M_{\rm WD} M_{\rm env}) / (4 \pi {R_{\rm WD}}^4)
\sim 2 \times 10^{19}~{\rm dyn~cm^{-2}}$,  
where $P^*$ is called the ``proper pressure,'' $M_{\rm WD}$ is the WD mass,
$M_{\rm env}$ the hydrogen-rich envelope mass, $R_{\rm WD}$ the WD radius.
Thus the ignition mass is independent of the mass-accretion rate.
This relation may originate from a formalism proposed by \citet{sha81}
and the numerically obtained critical pressures which were 
based on a simple envelope model of polytropic/adiabatic approximations
by \citet{mac83} and \citet{fuj82}.  This simple expression, however,
has discrepancy of envelope mass as large as an order of magnitude or more.
Numerical results obtained with non-adiabatic,
dynamical calculations with the OPAL opacity 
clearly show that the ignition condition is largely different from
the ``proper pressure'' and depends on the mass accretion rate
\citep[e.g.][]{pri95}.

\subsection{Chemical composition of ejecta}

Classical novae show heavy element enrichment in their ejecta
by an amount of a few to several tens percents by mass
\citep[e.g.,][]{geh98}.
The enrichment is interpreted as dredge-up of WD core material into
ejecta:  Hydrogen diffuses into WD cores during the accreting phase 
and ignites below the original WD core surfaces, thus a part of core 
material is mixed by convection into the hydrogen-rich envelope
and blown off in ejecta \citep{pri86}.  The diffusion occurs
when $\dot M_{\rm acc}\lesssim 10^{-9}~M_\odot$ yr$^{-1}$,
i.e., in a long timescale of $M_{\rm env}/\dot M_{\rm acc}
\sim10^{-4}~M_\odot/\dot M_{\rm acc}\gtrsim 10^5$ yr for classical novae.
Therefore, the diffusion process is not effective in recurrent novae, 
in which the recurrence periods are as short as 10--100 yr. 
Thus, heavy element enhancement is not expected.   

Recurrent novae, however, possibly show hydrogen depletion. 
In the very beginning phase of a nova outburst, convection widely develops
and carries processed helium up into the envelope. 
This convective mixing reduces the hydrogen content
by 10\%--20\% by mass for massive WDs like in recurrent novae. 
Therefore, we expect the hydrogen content of $X\approx0.50$ and
the solar metallicity of $Z\approx0.02$ in envelopes of recurrent novae.

Because dredge-up of WD core material is not theoretically expected in 
recurrent novae, it is hard to know whether WDs in recurrent novae
are made of carbon and oxygen (CO) or of oxygen, neon, and magnesium (ONeMg).
This is important from the SN Ia progenitor point of view,
because only CO WDs can explode as an SN Ia while
ONeMg WDs will collapse to a neutron star \citep{nom91}.
In the SD scenario, recurrent novae
contain either a CO WD \citep[increased from
a initial WD mass of $\lesssim 1.07~M_\odot$, see e.g.,][for 
an upper limit of CO WD masses]{ume99} or an ONeMg WD. 
In the DD scenario, all of the massive WDs ($\gtrsim1.07~M_\odot$) 
would be an ONeMg WD, because WDs hardly grow in
mass during and after a common envelope phase.

Recently, \citet{mas11} claimed that U Sco harbors an ONeMg WD 
because of a high Ne/O line ratio in spectra of the 2010 outburst,
comparable to those in several neon novae, but much larger
than those in CO novae.  Unfortunately there are no such data
in other recurrent novae other than classical novae. 
As we show later in Section \ref{sec:SD}, recurrent novae
have a very different evolutional history from those of classical novae.
Mass-increasing WDs like in recurrent novae
develop a thin helium layer underneath
the hydrogen burning zone.  After the helium shell grows in mass to
reach a critical value, a helium shell flash occurs \citep[e.g.,][]{kat89}.
Helium burning produces a substantial amount of Ne and Mg \citep{sha94}.  
Even if the WD is made of carbon and oxygen, the hydrogen-rich
envelope could be contaminated with neon and magnesium.
Therefore, some recurrent novae
would show neon enhancement if it occurs shortly
after the latest helium shell flash, because helium accumulation 
after every recurrent nova outburst could hide or dilute the products of
helium shell flashes.
Large Ne/O line ratios of ejecta do not always indicate an ONeMg WD.
Instead, it may indicate that a CO WD develops a helium layer during every
recurrent nova outburst.
We suggest that the high Ne/O line ratio in U Sco \citep{mas11} is 
also an evidence of mass-increasing WD, although we need
more observational information on the chemical compositions of U Sco
and the other recurrent novae.

\subsection{Light curve analysis of recurrent novae:
how to determine white dwarf mass}\label{sec:lca}

After hydrogen burning sets in, the envelope rapidly changes
its structure from a geometrically thin shell to a bloated spherical
configuration.  As the envelope expands from the WD surface
to a red-giant size, the photospheric temperature rapidly decreases
down to $\log T$~(K)$  \sim 4.0$ or lower.
This spherically extended configuration is stable against hydrogen
shell burning.  Then the nuclear burning settles in a steady-state.
The envelope mass decreases owing to nuclear burning and 
wind mass-loss.  The hydrogen burning continues until the envelope mass
decreases to reach a critical value for the extinguish condition. 
In this decay phase, the photospheric radius gradually shrinks whereas
matter goes away.   
So the photospheric temperature increases with time \citep{kat94h,kat99}.
After the wind stops, the envelope mass still continues
to decrease owing to nuclear burning. This stage corresponds to a SSS phase 
because the temperature is high enough to emit supersoft X-rays. 
At the termination of nuclear burning, the envelope structure comes back
to a geometrically thin shell.

In the decay phase of nova outbursts, 
free-free emission of optically thin ejecta dominates continuum flux  
\citep[e.g.,][]{gal76}.  
\citet{hac06kb} modeled free-free emission light curves of novae
based on the optically thick wind model calculated by \citet{kat94h}.
The decline rate of the model light curves 
depends sensitively on the WD mass and weakly on the chemical composition 
of envelope \citep{hac06kb}. 
Especially, when the WD mass is close to the Chandrasekhar mass limit 
$M_{\rm Ch}\approx 1.4~M_\odot$, the light curve depends sharply
on the WD mass. 
This is because the WD radius is very sensitive to the increase in mass
near $M_{\rm Ch}$.  Therefore, we can in principle determine the WD mass
by fitting the model light curve with the observed one
\citep[see, e.g.,][]{hac10k}.

Here, we should note that dynamical calculations of nova outbursts 
have numerical difficulties, especially in the treatment of 
surface boundary condition. Nariai et al (1980) already pointed out that 
expansion velocities are very different among different groups 
(see their Table 4). We note that 
Israel code \citep{ida12} tends to derive violent outbursts with  
large expansion velocities and large mass-loss rates, 
due to inadequate boundary conditions 
instead of solving the envelope up to the photosphere. 
No dynamical code has been succeeded so far in calculating 
extended stages of nova outbursts and their light curves.
On the other hand, the optically thick wind theory was developed
in order to calculate such extended stage of nova outburst and
has been succeeded in reproducing a number of nova light curves
\citep{hac06a,hac06kb,hac07GKper,hac08,kat09v838her}
as well as expansion velocities \citep[see Figure 18 in][]{hac07GKper}.

One of the characteristic properties of recurrent nova light curves is a
mid-plateau phase, that lasts 10--60 days after the brightness
decayed by several magnitude. The plateau phase often overlaps 
with a SSS phase, i.e., it begins/ends when the supersoft X-ray flux 
rises/decays \citep[e.g.,][]{hac07kl, osb11, nes12}.
Such plateau phases can be explained by the contribution of
an irradiated disk as demonstrated by \citet{hkkm00} for U~Sco and
by \citet{hac06a,hac07kl} for RS~Oph. 
The presence of an accretion disk is also suggested in T CrB, in which   
the secondary maximum in its light curve can be interpreted
as the contribution from an irradiated large tilting-disk \citep{hac99k}. 
On the other hand, in classical novae, such mid-plateau phases are rarely
reported.  The contribution of disks is also suggested in classical novae,
but it appears only in a very late phase
of nova outbursts \citep[see, e.g.,][for V1494 Aql]{hac04kk}. 
It is because the orbital period is usually much shorter in classical novae 
than in recurrent novae, and thus classical novae have much smaller 
disks that do not contribute much to optical light curves.

We think that accretion disks are not entirely blown off during the outburst
in recurrent novae.  One of the indications is the presence of these
mid-plateau phases.  All the recurrent novae
shows more or less a plateau phase.
The other indications are the detection of
flickering 8 days after the U Sco 2010 eruption \citep{war10}
and formation of accretion disk at least 8 days after the 2010 outburst
of U Sco \citep{mas12}.
During the mid-plateau phase, orbital light curves of U Sco have
asymmetric multiple-peak shapes outside the eclipse \citep{sch12pl},
which can be reproduced by spiral arm structure on the accretion disk 
\citep[see, e.g.,][for V1494 Aql]{hac04kk}.
Therefore, it is very likely that the disk was not gone but
survived during the 2010 outburst.  On the other hand, some numerical
calculations indicated that the accretion disk is entirely disrupted
by the fast moving ejecta \citep{dra10}.  This seems to be inconsistent
with the presence of the disk only 8 days after the outburst.
We suppose that numerical calculation cannot resolve 
the accretion disk that has high density region near the equator.

Outburst light curves of recurrent novae are a summation of
each contribution from a WD, irradiated accretion disk,
and irradiated companion.  
Such composite light curves are modeled for RS Oph by \citet{hac06a},
U Sco by \citet{hkkm00}, T CrB by \citet{hac99k}, V394 CrA by \citet{hac00},
and CI Aql by \citet{hac01a}.  The resultant WD masses are listed in
Table \ref{tab:WDmass}.

\subsection{Maximum magnitude versus rate of decline relation}\label{sec:mmrd}

There is a statistical relation between the maximum $V$ magnitude,
$M_{V,\rm max}$, and the rate of decline, $t_2$ or $t_3$,
in optical light curves of classical novae \citep[e.g.,][]{sch57,del95}.
This is called the MMRD relation.  Distribution of individual 
novae in the $t_3$--$M_{V,\rm max}$ plane shows a main trend from
the left-top to the right-bottom but there is
a large scatter above/below the main trend \citep[e.g.,][]{dow00,kas11}.
\citet{hac10k} theoretically explained the main trend and scatter
of the MMRD relation.  The main parameter that determines the main trend
is the WD mass, and 
the second parameter for the scatter
is the ignition (initial envelope) mass, 
in other wards, the mass accretion rate to the WD.
If the mass-accretion rate to the WD is relatively larger, the ignition
mass is smaller (see Fig. \ref{fig:MwdMdot}),
so that the peak brightness is fainter.  This second parameter
can reasonably explain the scatter of individual novae
\citep[see, e.g., Fig. 15 of][]{hac10k}.

It is, however, frequently discussed that the MMRD relation cannot be
applied to recurrent novae.
Table \ref{tab:WDmass} show the difference between the absolute 
$V$ magnitude and what the MMRD predicts, that is,
$\Delta M_{V, \rm MMRD} = M_{V,\rm MMRD} - M_{V, \rm max}$,
for six recurrent novae and
three very fast classical novae.  Here we estimate the absolute
$V$ magnitude $M_{V, \rm max}$ using the distance and extinction
in Table \ref{tab:WDmass}.  For the MMRD relation, we use (1)
\citet{dow00}, that is, 
$M_{V,\rm MMRD}= -10.79\pm0.92+(1.53\pm1.15) \log t_2$ for faster novae
($\log t_2 \le 1.2$),
but $M_{V,\rm MMRD}= -8.71\pm0.82+(1.03\pm0.51) \log t_2$
for slower novae ($\log t_2 \ge 1.2$), and also, for comparison,
(2) \citet{del95}, that is, $M_{V,\rm MMRD}= 
-7.92-0.81\arctan((1.32-\log t_2)/0.23)$.
Four recurrent novae, CI Aql, RS Oph, IM Nor, and T Pyx show 
reasonable agreement (within 1 $\sigma$ error) with what two MMRD
relations (1) and (2) predict.  Two classical novae
V838~Her and V2491~Cyg also show good agreement with what MMRD
relation (2) predicts.

On the other hand, U~Sco and T~CrB have a very similar absolute $V$
magnitude at maximum, $M_{V, \rm max}=-7.5$ and $-7.6$, respectively,
and both of them show a large deviation from the MMRD prediction.
Even if we use MMRD relation (2), the absolute magnitude of
U~Sco is still 1.5 mag fainter than the prediction. 
The symbiotic classical nova V407 Cyg also shows a very large deviation
from the MMRD relation. 
We think that these large deviations from the MMRD relation
are one of the indications of super-Chandrasekhar mass WDs
(Section \ref{sec:superchandra}) because a large deviation means
a small envelope mass at the optical peak (i.e., small ignition mass),
which is a strong indication of very massive WDs as 
explained in section \ref{sec:instability}.

\begin{figure}
\centerline{\includegraphics[width=12cm]{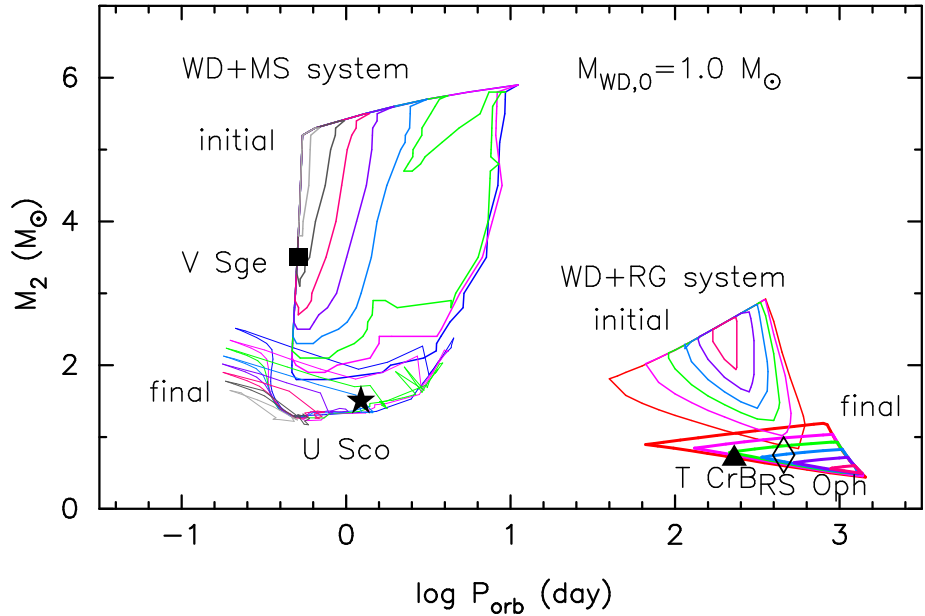}}
%
%

\caption{ Initial and final binary parameter ranges proposed
by a recent SD scenario \citep{hkn12b}.
Here, $M_2$ is the companion mass and $P_{\rm orb}$ the orbital period.
An initial binary system inside the region encircled by a solid line
(labeled ``initial'') is increasing its WD mass up to the
mass of $M_{\rm WD}= 1.38$, 1.5, 1.6, ..., 2.1, and $2.2~M_\odot$
(from outside to inside) and then reaches the regions labeled
``final'' when the WD stops growing in mass.
Currently known positions of recurrent novae
and supersoft X-ray sources are indicated by star mark ($\star$)
for U~Sco \citep[e.g.,][]{hkkm00}, filled square for V~Sge \citep{hac03kc},
filled triangle for T~CrB \citep[e.g.,][]{bel98}, and open diamond
for RS~Oph \citep[e.g.,][]{bra09}.
In the very last stage, the companion mass is reduced to
1--2~$M_\odot$ for the WD+MS systems, and $\lesssim1~M_\odot$
for the WD+RG systems.
\label{fig:SD}}
%
%
%
\end{figure}

\section{Evolutionary status of recurrent novae}\label{sec:SD}

In this section, we discuss the evolutionary status of recurrent nova
systems and how they will evolve toward SN~Ia explosions.
In the original DD scenario, there are no known paths to recurrent novae
especially for the RS~Oph type systems.  In the SD scenario, there are
two channels to SNe~Ia, i.e., the WD+MS (or simply MS) channel and WD+RG
(or symbiotic) channel. 
In the MS channel, a pair of MS stars evolve to a binary consisting 
of a helium star plus an MS star after the first common envelope evolution
\citep[e.g.,][]{hknu99}.  The primary helium star further evolves to
a helium RG with a CO core and fills its Roche lobe followed by
a stable mass-transfer from the primary helium RG to the secondary MS star.
The secondary MS star increases its mass and is contaminated by the primary's
nuclear burning products.  The primary becomes a CO WD after all the helium
envelope is transferred to the secondary MS star.  Then the secondary
evolves to fill its Roche lobe and mass transfer begins from the secondary
MS (or subgiant) star to the primary CO WD.
 When the mass-transfer rate exceeds
$\dot M_{\rm cr}$, an accretion wind evolution is realized.
The mass-transfer rate gradually decreases to below $\dot M_{\rm cr}$
and the optically thick winds stop.  
Then all the accreted matter is burned on the primary CO WD. 
The binary enters a persistent supersoft X-ray source (SSS) phase.
We should note that we must distinguish this persistent SSS phase
from transient SSS phases in classical novae and recurrent novae. 
The mass-transfer rate further decreases to below $\dot M_{\rm st}$
and the binary enters a recurrent nova phase.

In Fig. \ref{fig:SD}, binaries within the area labelled ``initial''
evolve downward and reaches the ``final'' area
when its mass increases to $M_{\rm WD} > 1.38~M_\odot$.  The supersoft
X-ray source V Sge and RX J0513.9-6951 are regarded as a binary
corresponding to the accretion wind phase \citep{hac03kb,hac03kc},
i.e., these binaries are now evolving toward SNe Ia. 
The present status of U Sco and other recurrent novae
in Table \ref{table.RN} corresponds to the final stage
of the above three phases (wind, SSS, and recurrent nova).    
If a WD explodes during the wind phase, the resultant SN Ia shows
some indications of the interaction between ejecta and circumstellar matter
(CSM) and the presence of hydrogen, because the WD winds and matter stripped
from the companion should form a rather massive CSM.
If it explode during the SSS or recurrent nova phase,
there are weak indications
of hydrogen because the secondary MS mass are $\sim 1$--$2~M_\odot$
and their size is compact. 
In multidimensional dynamical calculations of SN Ia explosion, however,
a very massive companion is often assumed 
\citep[e.g.][2 - 6 $M_\odot$ MS]{kas10} 
and they conclude that the SD scenario is inconsistent with no UV excesses
as well as no detections of hydrogen lines.
However, we think that these secondary masses are rather large and
not realistic. 

In recent SN Ia progenitor models of \citet{jus11}, \citet{dis11},
and \citet{hkn12b}, they suppose that a rapidly rotating WD does not
explodes as an SN~Ia even if it exceeds $M_{\rm WD}=1.38~M_\odot$,
because the central density is too low to ignite carbon.
They further assume that it takes a long time
before the SN~Ia explosion.  Once it exceeds $M_{\rm WD}=1.38~M_\odot$,
then the explosion is postponed until it spins down and the central 
density increases to reach a critical density for carbon ignition.
During such a long waiting time, the companion evolves 
off the main-sequence and become a WD or transfers its mass to
the primary WD and becomes a cataclysmic variable with a short
orbital period.  When the primary WD explodes as an SN~Ia,
neither hydrogen nor a bright ex-companion is left, but only a rather dark 
WD or a red dwarf ex-companion is left. 
\citet{hkn12b} showed that such a dark ex-companion is in 
a majority of SN Ia remnants coming from the MS channel.  

In the symbiotic channel we start from a very wide binary in which 
the primary evolves to an asymptotic giant branch (AGB) star.
The binaries undergo a common envelope-like evolution during the superwind
phase of the primary AGB star \citep[e.g.,][]{hkn99}.
The separation of the binary shrinks to the orbital period of 30 -- 800 days
and it becomes a pair of a CO WD and an MS star.
Then the secondary MS evolves to an RG with a helium core and
fills its Roche lobe.  Then the system enters an accretion wind evolution.
After that, the binary becomes a persistent SSS,
and evolves finally to a recurrent nova.  Depending on
the binary parameters, the WD mass reaches $M_{\rm WD}=1.38~M_\odot$ 
during one of the three phases (wind, SSS, and recurrent nova). 
One of such evolutionary paths is plotted in Fig. \ref{fig:MwdMdot}
by a red solid line.
The SMC symbiotic X-ray source SMC3 corresponds to the SSS phase 
and symbiotic recurrent novae like T CrB do the last phase.
If the WD explodes as an SN Ia during the first two phases, 
hydrogen should be detected in the SN~Ia explosion, because  
substantial amount of matter is stripped out from the RG 
\citep[see, e.g.,][for SMC3]{kat12b}.  
In the recurrent nova phase, however,
the companion has already substantially exhausted
its envelope through the long-lasted mass-transfer phase.
Its envelope mass may be small ($ \lesssim 0.5~M_\odot$), 
because they have developed a helium core.

In the spin-up/spin-down scenario of SNe Ia, the SN explosion could be
postponed until the secondary RG evolved off the red-giant branch and became 
a helium (or CO) WD.  These dark companion WDs could not be detected 
just before and after the SN~Ia explosion.
\citet{hkn12b} showed that in their rapidly spinning WD progenitor models, 
these dark companions are the majority of SN Ia progenitors.
Therefore, we cannot expect any indication of hydrogen or bright companions
in such SN~Ia explosions.

Observations of SNe Ia have provided the following
constraints on the nature of companion stars.
Some evidences support the SD model, such as the presence of circumstellar
matter \citep{pat07, ste11, fol12} and detections of hydrogen
in the circumstellar matter-ejecta interaction type
SNe (type Ia/IIn) like SN~2002ic \citep{ham03}, SN 1604 (Kepler's SNR)
\citep{chi12}, and PTF11kx \citep{dil12}.
On the other hand, there has been reported no direct indication of
the presence of companions, which is usually thought to be unfavorable
for the SD scenario, e.g.,
(1) the lack of companion stars in the images of
SN 2011fe \citep{liw11}, some SN Ia remnants (SNRs) \citep{sch12p},
SN 1572 (Tycho's SNR) \citep{ker09} and SN 1006 \citep{ker12},
(2) the lack of UV excesses
 of early-time light curves \citep{kas10}, and
(3) the lack of hydrogen features in the spectra \citep{leo07}.
Both (2) and (3) are expected from the collision between ejecta
and a companion.  
\citet{hkn12b} showed that their SD scenario mainly produces unseen
companions at SN Ia explosion because of a long spin-down time.
They also showed that the WD mass distribution at SN Ia explosion
predicted by their model agrees well with the distribution of
SN Ia brightness observed if the brightness is correlated to 
the WD mass.  These results strongly support that a major route to
SNe~Ia is the SD system and they explode through a recurrent nova stage.

\section{Elementary processes in binary evolution}
\label{sec:elementary}

The SD scenario is based on different elementary processes from those 
assumed in the original DD scenario, which result in a very different binary
evolution as well as population synthesis.  Thus, it is important
to clarify theoretical backgrounds of each elementary process.
In this section, we critically examine each elementary process
based on the results of nova studies, because many of these
ingredients are closely related to nova phenomena.

%
%

\subsection{Common Envelope Evolution}\label{sec:CEE}

Common envelope evolution is a fundamental process of binary evolution
for both the SD and DD scenarios.  The first common envelope evolution
in a binary occurs when the primary star evolves to a red-giant and fills
its Roche lobe.  Common envelope evolutions were formulated
in a very simple form based on the total energy conservation
(gravitational energy of the orbit plus envelope binding energy)
before and after the common envelope evolution.
The ratio of the binary orbit before ({\it initial}) and after ({\it final}),
$A_f/A_i$, is given by a simple form using an efficiency parameter
$\alpha (< 1)$ \citep[e.g.][]{ibe93}. 
A smaller value of $\alpha$ leads to a smaller ratio of $A_f/A_i$, i.e.,
a larger shrinkage of the binary orbit.
In the original $\alpha$-formalism, $\alpha$ is only the parameter.
Some authors had a different expression to calculate the gravitational
energy of the RG envelope by using $\lambda$-parameter \citep{web84,dek90},
which represents the effective mass-weighted radius of RG stars.
Then the product of $\alpha\lambda$ behaves like the original $\alpha$.
Observationally, \citet{zor10} derived $\alpha = 0.2$--0.3, assuming
that $\alpha$ is universal together with $\lambda=0.5$,
while \citet{reb12} obtained $\alpha \sim 0.25$.
\citet{dem11} obtained the dependence of $\alpha$ on 
the mass ratio $q=M_2/M_1$ such as $\alpha \sim 1$ for $q=0.1$,
$\alpha \sim 0.2$ for $q=0.3$, and  $\alpha \lesssim 0.1$ for $q=1$.
These new $\alpha$-parameters work  
together with the effective radius parameter of $\lambda$, for example,
$\lambda=0.4$ for $M_1=5~M_\odot$ or $\lambda=0.45$ 
for $M_1=7~M_\odot$ of AGB stars \citep[e.g.,][]{dem11}.
These recent analyses suggest that $\alpha\lambda\sim0.1$--0.2. 
In many population synthesis calculations that support the DD scenario,
however, a larger value of $\alpha\lambda \sim1$ was preferentially
used independently of $q$, because such a larger value yields
a larger birth rate of SNe Ia in their DD models \citep[e.g.,][]{rui09}.
It is however interesting that the birth rate of SNe Ia
coming from SD binaries increases and
becomes comparable to the rate coming from DD binaries if we have
a smaller value of $\alpha\lambda\lesssim 0.5$ \citep[e.g.,][]{rui09}. 
This suggests that if we adopt a realistic value of $\alpha\lambda$
such as $\alpha\lambda\sim0.1$, the SD binaries dominate the progenitors
of SNe Ia in population synthesis.   

One of the recent trends in the population synthesis community
is to assume an extremely large value of $\alpha~(> 1)$
\citep{hur02, tou05, men12}. 
If there is a large additional energy source (e.g.,
recombination energy), the binary separation will even increase
after the common envelope evolution \citep{web08}.
However, such a mechanism seems to be very difficult to work effectively
because it needs all the envelope matter instantaneously recombined
in the accelerating zone and the released energy efficiently converted to
the binding energy without radiative loss
\citep[see, e.g., a criticism by][]{sok03}.
This idea may be introduced in order to reproduce symbiotic 
recurrent nova systems in the DD scenario \citep{web08}.
However, recent observational estimates,
$\alpha\lambda\sim 0.1$ for AGB stars, do not support such
an extremely high value of $\alpha\lambda$ $(>1)$ \citep[e.g.,][]{zor10}.

\citet{nel00} proposed a new formalism for common envelope evolution
based on the assumption that angular momentum loss is proportional to
the amount of lost mass, $\Delta J/J = \gamma \Delta M/M$.
This formalism contains a proportionality parameter $\gamma$, so 
dubbed ``$\gamma$-formalism.''
This formalism is criticized by \citet{web08} as it does not actually
limit $A_f$. \citet{zor10} also criticized the $\gamma$-formalism
because the $\alpha$-formalism is much more reasonable from
the statistics of a large number of binary parameters.

\subsection{Accretion wind evolution versus common envelope
evolution}\label{sec:EP}

After the first common envelope evolution,
the binary becomes a pair of a WD and a secondary star.
After some time elapses, the secondary evolves to fill its Roche lobe.
Then the mass transfer begins from the secondary to the primary WD
through the L1 point.  Infalling matter forms
an accretion disk and finally accretes on to the WD.
The accreted matter quickly spread over the entire
WD surface and hydrogen nuclear burning ignites after some critical 
amount of mass accumulates. 
If the secondary is an evolved RG more massive than the WD,
the mass accretion rate exceeds the critical value,
$\dot M_{\rm cr}$ (see Fig. \ref{fig:MwdMdot}), and
strong optically thick winds blow.  
Once the wind occurs, the wind carries away mass and angular momentum.
Then the orbital separation and the size of the secondary's Roche lobe
change in time.  The mass-transfer rate is regulated to keep the
secondary's radius equal to the Roche lobe radius.  If the mass-transfer
rate is smaller than  $\dot M_{\rm acc}\lesssim10^{-4}~M_\odot$ yr$^{-1}$,
a common envelope does not form \citep{hkn99,hknu99}.
This is the accretion wind evolution (Fig. \ref{fig:accretionwind}).
When the accretion rate is very large
$\dot M_{\rm acc}\gtrsim10^{-4}~M_\odot$ yr$^{-1}$, however, 
the photospheric radius of the WD expands
over the Roche lobe, and a common envelope evolution may occur.

If the accretion wind evolution is taken into account, the condition for
the second common envelope evolution is relaxed to $q < 1.15$ from
$q < 0.79$, where $q=M_{\rm RG}/M_{\rm WD}$. The former condition is
obtained with the mass and angular momentum conservations including
those of the winds, and the latter without the winds.
Therefore, some binaries can avoid a formation of the second common envelope
and keep the binary separation.  The binary still consists of
a mass-accreting WD and a lobe-filling secondary.
The WD can grow in mass during the accretion wind evolution \citep{hkn96}.

This is the first step that widens the binary parameter range of SN Ia
progenitors.  It is interesting to note that optically thick winds
have been established in nova outbursts. 
The U Sco 2010 outburst suggested that an accretion disk
formed (or still existed) in a very early phase, only
8 days after the optical maximum 
\citep[see Section \ref{sec:lca} and][]{mas12}. 
In this early phase, the optically thick wind had not yet stopped
(it ends at around Day 13) but the companion and accretion disk emerged from
the extended envelope at Day 8 \citep{hac12kc}.
This means that accretion from the companion can be observed
when the companion emerged from the WD photosphere,
while the optically thick wind still continues.
We think that this is an observational example of accretion wind evolution.

\subsection{Mass stripping in accretion wind evolution}\label{sec:stripping}

When the accretion wind occurs as illustrated in
Fig. \ref{fig:accretionwind}, the fast wind ($\gtrsim 1000$ km~s$^{-1}$)
hits the companion's surface that fills the Roche lobe.
A very surface layer of the companion's hemisphere may be stripped
by the winds, which reduces the mass outflow rate from the secondary.
\citet{hkn99} first incorporated this effect into their binary evolution
and established the symbiotic channel to SNe Ia.
This is the second step that widens the parameter range of
binary evolution toward SNe Ia.

The accretion wind evolution including the mass-stripping
effect is also essential to understand the mechanism of quasi-periodic
behavior of the two SSS binaries, V Sge and RX J0513.9$-6951$. 
\citet{hac03kb,hac03kc} explained both the optical 
and supersoft X-ray light curves of these two objects based on
the accretion wind evolution that leads to a limit cycle of the
light curve behavior.  They obtained the WD masses of
$\sim 1.2$ -- 1.3 $~M_\odot$, and the mass-increasing rates of the WDs
$\sim 10^{-6}~M_\odot$ yr$^{-1}$ for the both objects from light curve
fitting.  Therefore, these two systems are also candidates
of SN Ia progenitors.
There are several similar objects in our Galaxy, the so-called
V Sge-type stars \citep{ste98}.   \citet{kaf08} and \citet{lin08}
found that QU Car is a very similar object to V Sge and belongs to
the V Sge-type stars.  This also suggests that many such objects
could be misclassified to other categories of variable stars and
the number of V Sge-type stars would significantly increase. 

The successful results of limit-cycle light curves strongly
suggest that (1) both the processes of ``accretion wind'' and
``mass stripping'' can be realized in binary evolution 
and (2) we expect a wider parameter region of binary evolution
toward SNe Ia by avoiding the second common envelope evolution.

\subsection{AGB super winds in wide binaries}\label{sec:agbwind}

Stars with a mass less massive than $M\lesssim8~M_\odot$ lose
most of its envelope mass in a short timescale ($\lesssim10^4$~yr)
at the final stage (at its asymptotic giant branch, AGB, phase)
of evolution.   \citet{hkn99} incorporated this AGB superwind
as an important process in binary evolution.
Due to large mass-loss rates of superwinds \citep[up to several
$\times10^{-4}~M_\odot$ yr$^{-1}$,][]{gro02,gua06}, even
very wide binaries ($a_i=1500$ -- 30000 $R_\odot$) experience
a decay of the orbital separation that is similar to a common envelope
evolution.  After that, it becomes a wide binary
corresponding to the orbital periods of symbiotic stars.
This is the symbiotic channel of the SD scenario first
proposed by \citet{hkn99}.

The original DD scenario did not include the evolution driven by
superwinds in very wide binaries.  
Many population synthesis calculations still started from
zero-age binaries excluding very wide binaries.  
Thus, after the first common envelope evolution, the separation shrinks
from $a_i \sim 700$--2200$~R_\odot$ to
$a_f \sim 5$--$17~R_\odot$ ($P_{\rm orb}\sim 0.5$--1.5 days)
if we use $\alpha\lambda\sim0.1$ from
the recent trend of small values of $\alpha$ (see Section \ref{sec:CEE}).
Then the resultant binary periods lie in the region of 0.5--1.5 days,
which is not consistent with the present states of symbiotic recurrent novae
but consistent with the MS channel of SD scenario.
This is the reason why the SN Ia rate coming from SD binaries
becomes large and exceeds the rate coming from DD binaries
if we assume a small value of $\alpha\lambda$. 


\subsection{Mass increasing rate of white dwarfs}\label{sec:Hflash}

The original DD scenario was based on the mass increasing efficiency
of WDs being very different from that of the SD scenario. 
In the original DD scenario, mass-accreting WDs can grow in mass
only if it stays in the narrow strip region 
($\dot M_{\rm st} \le \dot M_{\rm acc} \le \dot M_{\rm cr}$)
in Fig. \ref{fig:MwdMdot}.
These WDs, however, suffer helium shell flashes, which
take away a large part of the helium envelope mass. 
Thus the total efficiency of mass-increase is very small. 
Such a small mass-increasing efficiency is based on the claim by
\citet{cassi98} and \citet{pie99,pie00} who calculated shell flashes on
low-mass WDs ($\lesssim0.8 M_\odot$) using a spherical symmetric hydrostatic
code with the Los Alamos opacity. They concluded that
the frictional process due to the companion's motion should be
very effective to take away most of the envelope mass,
simply because the envelope photosphere expands beyond the binary orbit,
although their calculation did not include any frictional effects
in their codes.

It is worth noting that \citet{cassi98} emphasized 
'the Los Alamos opacities are very similar to the OPAL opacities'
(see the last sentence of Section 4 in their paper).
However, this is clearly an incorrect statement 
\citep[see a comparison between nova envelopes calculated with the new
and old opacities in Fig. 15 of][]{kat94h}.
As we have already known,
calculations with the old opacity could derive very
different conclusions.  Even in the same framework of 1980's physics,
however, their claim was not supported by the results given by
\citet{kat89} and \citet{kat91a,kat91b}.
Now a days we have many calculational/observational indications that
frictional processes are ineffective in nova outbursts
[see, e.g., introduction in \citet{kat11}, 
Section 7.2 in \citet{kat12}, and also
Section \ref{sec:v445pup} of this review].
\citet{cassi98} and \citet{pie99,pie00} extended their arguments
on the low mass WDs ($\lesssim0.8 M_\odot$)
to all the WD masses up to the Chandrasekhar mass,
as often cited like 'WDs are hardly grow to reach
the Chandrasekhar mass limit due to frictional effects' 
\citep[e.g.][]{ste03,yun05,mao10}
in the arguments against the SD scenario.

In the SD scenario, on the other hand, we have a much wider mass-growing
region of WDs,  not restricted into the narrow strip
in Fig. \ref{fig:MwdMdot} as in the original DD scenario.
In addition to this narrow strip,
WDs in the region above $\dot M_{\rm acc} = \dot M_{\rm cr}$ can grow in mass
during the accretion wind evolution.  It could also grow in the region
slightly below the line $\dot M_{\rm acc} = \dot M_{\rm st}$, i.e.,
$10^{-8}~M_\odot$ yr$^{-1}\lesssim\dot M_{\rm acc}<\dot M_{\rm st}$,
where the shell flashes are weak.  The mass increasing efficiency
of WDs is obtained from the combination 
($\eta_{\rm H}\eta_{\rm He}$) of the mass accumulation efficiencies
of hydrogen shell burning/flashes $\eta_{\rm H}$ \citep{hknu99, pri95}
and of helium shell burning/flashes $\eta_{\rm He}$ \citep{kat04}.

\citet{nel12} compared the total mass-increasing efficiency
of a 1 $M_\odot$ WD between the DD and SD models.
The mass increasing efficiency is up to 
$\eta_{\rm H}\eta_{\rm He}\sim 0.8$ in the SD models,
but only $\eta_{\rm H}\eta_{\rm He}\sim 0.1$
at most (at $\log \dot M_{\rm acc}~(M_\odot~{\rm yr}^{-1}) \sim -6.6$)
in the model by Yungelson (2010) \citep[see Fig. 4 in][]{nel12}.
The latter value is increased to $\eta_{\rm H}\eta_{\rm He}\sim0.2$
 (at $\log \dot M_{\rm acc} \sim -6.1$) for
symbiotic stars \citep{ibe96} because of their larger Roche lobe sizes.
Now we can see where the difference comes from.
The original DD model \citep{ibe96} adopt the old opacity
and instantaneous envelope mass ejection due to the Roche lobe overflow.
Moreover, they assumed line-driven winds from the WD of very large 
mass-loss rate (with 100\% efficiency of photon momentum flux,
i.e., a few times $10^{-7}M_\odot$ yr$^{-1}$).  We think that this
rate is too large and unlikely.
Such large mass-loss rates are not included in the SD scenario because
no optically thick winds blow when $\dot M_{\rm acc}< \dot M_{\rm cr}$.
As a results, the original DD scenario yields a very small efficiency
of mass-growth of WDs even when they calculate SD binaries
in their DD scenarios.  Thus, the SD channel hardly contributes to 
the total SN Ia rate in the DD scenarios.

\subsection{Other ingredients -- accretion disk instability}\label{sec:king}

\citet{kin03} presented an idea that thermal disk instabilities
temporarily increase the mass accretion rate onto the WD,
which would trigger steady hydrogen burning,
thus, the lower line of the narrow strip ``steady H-burning''
in Fig. \ref{fig:MwdMdot} would go down as much as by $-2.4$
in logarithmic scale.  However, this idea is based on a misunderstanding
about the instability of hydrogen shell-burning (shell-flash). 
We should note that the same mistake was repeated in \citet{ale11} after
the criticism  by \citet{hkn10}.
As explained in Section \ref{sec:instability},
a shell flash occurs when the envelope mass increases to the
ignition mass at which the thermal balance in the envelope
breaks down.  This does not occur only when the accretion rate
temporarily exceeds $\dot M_{\rm cr}$.
Unfortunately, this incorrect idea was already incorporated
in some binary evolution models
\citep{xu09,wan10,men10}, but these results have no astrophysical meaning.

\section{Related objects}\label{sec:relatedobj}

\subsection{Super-Chandrasekhar mass white dwarfs
 -- U Sco, T CrB, and V407 Cyg}\label{sec:superchandra}

Recently super-Chandrasekhar mass WDs are suggested from
observations of very luminous SNe~Ia \citep[e.g.,][]{how06, hic07, yam09,
sca10, sil11, tau11}.  In the SD scenario,
WDs that exceed the Chandrasekhar mass limit
are theoretically possible if they are rapidly rotating
\citep[e.g.,][and references therein]{jus11, dis11, hksn12a, hkn12b}.
If recurrent novae are progenitors of SNe Ia, some of them could be
super-Chandrasekhar mass WDs.
So far, only one object (U Sco) was suggested
to harbor a super-Chandrasekhar mass WD \citep{hksn12a},
but we here suggest two more objects, the symbiotic recurrent nova T~CrB
and symbiotic classical nova V407~Cyg.

The WD mass of U Sco was observationally estimated to be
$M_{\rm WD}= 1.55 \pm 0.24 ~M_\odot$ in the 1999 outburst \citep{tho01}.
This value is very suggestive, although the ambiguity is too large
to draw a definite conclusion of super-Chandrasekhar mass.
There are additional indications of super-Chandrasekhar mass WD in U Sco.
(1) The present binary parameters of U Sco shows that it just locates
in the ``final'' stage of the path toward super-Chandrasekhar mass WDs 
in the SD scenario (Fig. \ref{fig:SD}),
(2) The very fast decline of the optical nova light curve is  
consistent with a very massive WD as large as (or beyond)
the Chandrasekhar mass limit of no rotation ($M_{\rm Ch}=1.4~M_\odot$).
(3) The maximum brightness, $M_{V, \rm max}=-7.5$, is as much as 3.0 mag below 
the MMRD's prediction (Section \ref{sec:mmrd}).  There is another candidate
of super-Chandrasekhar mass WD, the symbiotic recurrent nova T CrB.  
The early decline of its light curve is so rapid and its maximum brightness
shows a large deviation (2.4 mag) from the MMRD relation
(see Section \ref{sec:mmrd}).
We think this kind of large deviations ($\gtrsim 2.5$ mag, denoted
by asterisk in Table \ref{tab:WDmass}) from the
MMRD relation is an indication of super-Chandrasekhar mass WD unless
the estimated distance is largely different from the true one.

V407~Cygni is a long orbital period (D-type)
symbiotic star consisting of a WD and a Mira \citep{mun12}.
V407~Cyg outbursted as a classical nova in 2010 \citep{nis10}.
The optical light curve shows a very rapid decline in the very early phase
which resembles that of U Sco.  It entered a plateau phase
only several days after the outburst and started
the final decline 47 days after the outburst \citep{mun12}.
Such an evolution of light curve is faster than that of RS~Oph.
\citet{hac12kc} modeled the optical light curve of V407 Cyg and
suggested that the WD mass of V407 Cyg as massive as that of U Sco and
its plateau phase can be reproduced by a large irradiated disk like in 
RS~Oph \citep{hac06a}.
The peak absolute magnitude is calculated to be $M_{V,\rm max}=-6.6$
as in Table \ref{tab:WDmass} and the MMRD error
is as large as 3.0 mag almost similar to that of U Sco.
Thus, we expect that V407 Cyg is also a candidate for
super-Chandrasekhar mass WD.

Ejecta of the V407 Cyg 2010 nova outburst were enriched with heavy elements
\citep{sho11}, suggesting that WD core material was dredged up and
ejected.  This means that the WD mass decreases after every nova outburst.
Thus, V407~Cyg will possibly not explode as an SN Ia if the WD has a
sub-Chandrasekhar mass.  If the WD is composed of carbon and oxygen
and has a super-Chandrasekhar mass,
however, it could be a progenitor of SN~Ia because even after many nova
cycles the WD is still more massive than the Chandrasekhar mass
and can explode soon after it spins down \citep{hkn12b}.
This kind of symbiotic progenitors can explain the interaction
between SN~Ia ejecta and circumstellar matter as seen
in Kepler's supernova remnant (SNR) \citep{rey07,chi12} and in PTF~11kx
\citep{dil12}.

\subsection{Helium nova V445 Pup}\label{sec:v445pup}

V445 Pup is a helium nova that underwent a helium shell flash
\citep{kam02,kat03,ash03,wou09,kat08}.
The WD mass is estimated to be very massive $M_{\rm WD}\gtrsim1.35~M_\odot$,
and the companion star is a slightly evolved helium star of mass 
$\gtrsim0.8~M_\odot$ \citep{kat08}.
The orbital period is suggested to be $P_{\rm orb}=0.65$ days \citep{gor10}.
The WD is already as massive as (or close to) the Chandrasekhar mass
and its mass is increasing after one cycle of helium nova.
Therefore, this system is a candidate for SN Ia progenitors \citep{kat08}.
Such a binary, however, does not fit to any known path
in the binary evolution scenarios \citep[for example, the final orbital
periods are much shorter than 0.65 days in their helium star donor
channel of][]{wan09,wan10b}, indicating the unknown
third path toward SNe Ia or some missing processes
in binary evolution models of helium star donor channel.

In mass increasing WDs like in SSSs or in recurrent novae, we expect that a
helium layer develops underneath the hydrogen burning zone.
This helium layer periodically experiences helium shell flashes.
V445 Pup indicates that some accreting WDs can become very massive
and neither helium nor carbon detonation has ever occurred
at the bottom of the helium layer. This does not support the idea
that accreting WDs cannot grow in mass due to frictional process,
which is frequently assumed in the DD scenarios in order to
reject the SD scenarios.

\section*{Acknowledgements}

This research has been supported in part by the
Grant-in-Aid for Scientific Research (22540254 and 24540227)
of the Japan Society for the Promotion of Science. 


\label{lastpage}
\end{document}